\documentclass[12pt]{iopart}

\usepackage{epsfig,psfig,amssymb}

\begin{document}

\title[One-dimensional non-interacting fermions in harmonic confinement]{One-dimensional non-interacting fermions in harmonic confinement:
equilibrium and dynamical properties}

\author{Patrizia Vignolo\footnote[3]{To
whom correspondence should be addressed (vignolo@sns.it)} and Anna Minguzzi}
\address{NEST-INFM and Scuola Normale Superiore,
Piazza dei Cavalieri 7, I-56126 Pisa, Italy}

\begin{abstract}
We consider a system of one-dimensional non-interacting fermions in
external harmonic confinement. Using an efficient Green's function
method  we evaluate the exact profiles and the pair correlation
function, showing a direct signature of the Fermi statistics and of
the single quantum-level occupancy. We also study the dynamical
properties of the gas, obtaining the spectrum both in the
collisionless and in the collisional regime. Our results apply as well
to describe a one-dimensional Bose gas with point-like hard-core
interactions.
\end{abstract}

\maketitle

\section{Introduction} 
Atomic Fermi vapours have been cooled down to degeneracy in several
experiments on ultra-cold  atoms of $^6$Li and $^{40}$K
\cite{jila_exp,rice_exp,paris_exp}, using  techniques similar to those
employed to obtain Bose-Einstein condensation \cite{bec}. At very low
temperatures collisions between atoms are essentially in the $s$-wave channel;
however, all even-channel collisions  are forbidden for spin-polarized
fermions by the Pauli exclusion principle. Since
dipolar interactions are also negligible, the Fermi
gas can be considered as non-interacting. 
In the experiments the atoms are trapped by magnetic or optical
potentials, providing a confinement which can be well approximated as
harmonic. The geometry of the trapping potential can be very
anisotropic, thus allowing to reach quasi-1D or quasi-2D situations
\cite{ketterle_1d}. 

We focus here on the quasi-1D geometry, which is the case of the Paris
experiment \cite{paris_exp} and is relevant for experiments on atom
lasers and atomic waveguides. The 1D gas of non-interacting fermions
serves also as a model for describing the 1D Bose gas with hard-core
point-like interactions (Tonks gas, \cite{tonks}).
 Indeed, it was demonstrated by
Girardeau that there exists an exact mapping between
the bosonic and 
the fermionc many-body wave functions, $\Psi_B=|\Psi_F|$, valid also
in the time domain \cite{girardeau,girardeau_td}. The Tonks gas limit
corresponds to the case of total reflection in a two-atom collision
\cite{olshanii}, 
and the interactions in 1D mimic the role of the Fermi statistics.

In this paper we present some results for the exact equilibrium
profiles and for the dynamical properties of a 1D Fermi gas in
harmonic confinement. In Sec.~\ref{sec2} we illustrate the role of
quantum statistics on the particle and kinetic-energy density
profiles, and we investigate how the Pauli principle influences the
pair correlation function. In Sec.~\ref{sec3} we evaluate the spectrum
of the gas both in the collisionless and in the collisional
regime. In addition, we relate the result for the frequency of the
quadrupole mode to a scaling property of the Hamiltonian used in the
long-wavelength limit. Finally,  Sec.~\ref{sec4} offers a
summary and some concluding remarks.

\section{Equilibrium properties}
\label{sec2}

We consider a system of $N$ non-interacting Fermi particles confined
in an external potential which is very tight in two spatial directions.
We assume that the gas is in a quasi-1D
situation where the transverse degrees of freedom are frozen, that is,
the chemical potential is much smaller than the level spacing in the
transverse direction.
In this case the transverse wave function is the ground state one,
which will factorize out from all the following
calculations. 
We can thus focus only on the axial coordinate $x$ and define
the external confinement along $x$ given by a real
potential $V_{ext}(x)$. 
For the sake of simplicity we will assume that the
eigenfunctions $\psi_i(x)$ are real~\cite{noticina}.

The equilibrium properties of the gas can be expressed in terms of the  
longitudinal one-body density matrix, which at zero temperature reads
\begin{equation}
\rho^{(l)}(x,x')=\sum_{i=1}^N\psi_i(x)\psi_i(x').
\label{rho_def}
\end{equation}
In the following we will give several examples of calculations of the
exact spatial profiles obtained as the moments of the density matrix, and 
we will evaluate also the longitudinal pair distribution function. 

\subsection{Density profiles}
\label{density_prof}
The  zeroth moment of the one-body density matrix yields the
equilibrium density profile $n(x)$ 
\begin{equation}
n(x)=\rho^{(l)}(x,x')|_{x'=x}=\sum_i\langle\psi_i\,|\delta(x-x_i)|\,\psi_i\rangle,
\end{equation}
where $x_i$ are the positions of the fermions and $|\,\psi_i\rangle$ are 
the eigenstates of the Hamiltonian in the coordinate representation.
We have shown~\cite{noi} that $n(x)$ can be written in terms of the Green's
function in coordinate space, $\hat G(x)=(x-\hat{x}+ i \varepsilon)^{-1}$,
which is a function of the position operator $\hat x$:
\begin{equation}
n(x)=-\frac{1}{\pi}\lim_{\varepsilon\rightarrow0^+} {\rm
Im}\,\sum_{i=1}^N \langle \psi_i\,|\hat G(x) |\,\psi_i\rangle\;.
\end{equation}
This equation expresses the density profile as   
the trace of the operator $\hat G(x)$ on the first $N$ quantum 
states (${\rm Tr}_N$). 
Using the relation between the partial trace ${\rm Tr}_N$ of a generic matrix 
$Q$  and the determinant of the inverse matrix $Q^{-1}$,
\begin{equation}
{\rm Tr}_N Q=\partial[\ln\det(Q^{-1}+\lambda{\Bbb{I}}_N)]/
\partial\lambda|_{\lambda=0},
\label{anna}
\end{equation}
we obtain 
\begin{equation}
n(x)=-\frac{1}{\pi}\lim_{\varepsilon\rightarrow0^+} {\rm
Im}\,\frac{\partial}{\partial\lambda}
[\ln\det(x+i\varepsilon-\hat x-\lambda{\Bbb{I}}_N)]_{\lambda=0}\;.
\end{equation}
This alternative expression for the density profile is very practical
for performing numerical calculations in the case of harmonic
confinement, $V_{ext}(x)=m\omega_{ho}^2x^2/2$. 
For this specific potential the representation for both
the position and the momentum operators 
on the basis $\psi_i$ of the eigenstates of the harmonic oscillator
is a tridiagonal matrix. Thus the operator
$x+i\varepsilon-\lambda{\Bbb{I}}_N$ 
is tridiagonal: the determinant of such a matrix can be 
evaluated recursively
through a renormalization procedure~\cite{noi,papirone1}.

\begin{figure}
\centerline{\epsfig{file=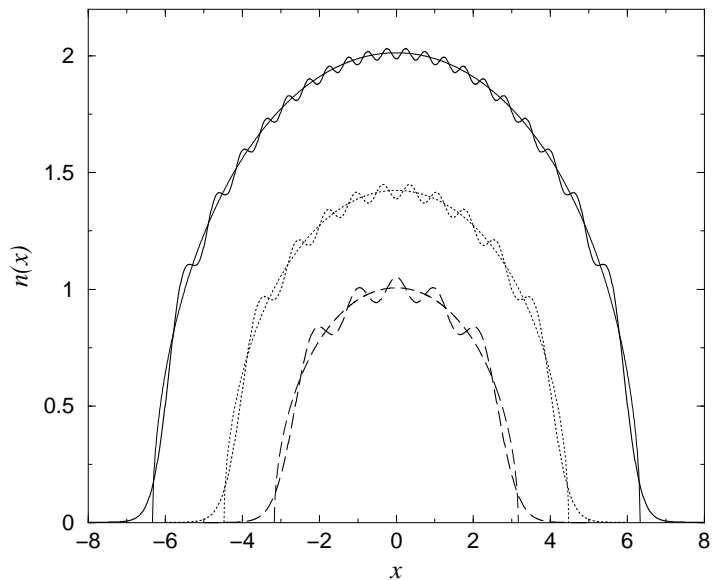,width=0.6\linewidth}}
\caption{Exact particle density $n(x)$ (bold lines) and local density
approximation (thin lines) in units of $a_{ho}^{-1}$
as functions of the spatial coordinate $x$ (in units of 
$a_{ho}=\sqrt{\hbar/m\omega}$) for $N$=5 fermions (dashed line), 
10 fermions (dotted line) and 20
fermions (solid line).}
\label{fig1}
\end{figure}

The numerical results for the particle density profiles are shown in
Fig.~\ref{fig1} at various number of fermions, as compared  with the
predictions of the local-density approximation: the latter neglects
the shell structure and the spill-out from the classical border. 
Because of the symmetry between position $\hat x$ and momentum $\hat p$ in the
harmonic-oscillator Hamiltonian, the momentum distribution has 
exactly the same behaviour as the particle density illustrated in
Fig. \ref{fig1}. 

On account of the mapping with the Tonks gas the particle density
profile of $N$ noninteracting fermions in the 1D harmonic trap is identical
to that of $N$ bosons with hard-core interactions in the 
same external potential, while the momentum distribution for
such a system has a sharp peak at small wavevectors~\cite{Girardeau_write}.

\subsection{Kinetic energy and momentum flux densities}
The first moment and all the other odd moments of the one body density
matrix are zero because of the reality of the wave functions.
For higher even moments several definitions are avaible in the literature
(for a review see {\it e.g.} Ref.~\cite{uhlenbeck}); these  can be 
expressed again in terms of $\hat G(x)$~\cite{papirone1}.
In particular, we shall concentrate on the quantity  
\begin{eqnarray}
P_n(x)&=&(-i)^n\left[\frac{\partial^n}{\partial
{x'}^n}\rho^{(l)}(x,x')\right]_{x'=x}\nonumber\\
&=&-\frac{1}{\pi}\lim_{\varepsilon\rightarrow0^+} {\rm
Im}\,\sum_{i=1}^N \langle \psi_i\,|\hat G(x)\hat p^n |\,\psi_i\rangle\; 
\label{stella}
\end{eqnarray}
and on its symmetrized form
\begin{eqnarray}
S_n(x)&=&(-i)^n\left[\frac{\partial^n}{\partial
r^n}\rho^{(l)}(x+r/2,x-r/2)\right]_{r=0}\nonumber\\
&=&-\frac{1}{\pi}\frac{1}{2^n}\sum_{\sigma=0}^n\left(\begin{array}{c}
n\\\sigma\\\end{array}\right)\lim_{\varepsilon\rightarrow0^+}
{\rm Im}\,\sum_{i=1}^N \langle \psi_i\,|\hat{p}^{\sigma}\hat G(x)\hat
p^{n-\sigma} |\,\psi_i\rangle\;. 
\label{stellastella}
\end{eqnarray}
Here we have set $\hbar=1$ and $m=1$.
For $n=2$ these definitions give rise respectively to twice the
local kinetic 
energy $T(x)$ [Eq. (\ref{stella})] and to the momentum flux 
density $\Pi(x)$ [Eq. (\ref{stellastella})]. 
Although the semiclassical limits and the integrals of these two
functions  coincide, 
in the quantum limit  these are 
different functions: the kinetic energy density is invoked in
density-functional theory, while the momentum flux density appears in
the equations of generalized hydrodynamics.

The same method which was used in Sec.~\ref{density_prof} 
to express the density profile
in terms of the determinant of the inverse of the Green's function
operator can be extended to deal
with the operators of the form $\hat p^\sigma \hat G(x) \hat
p^{n-\sigma}$ \cite{papirone1}.
Again, for low values of $n$ one is reduced to
evaluate with recursive methods the determinant of sparse matrices, which are
always tridiagonal in the tails. 

\begin{figure}
\centerline{\epsfig{file=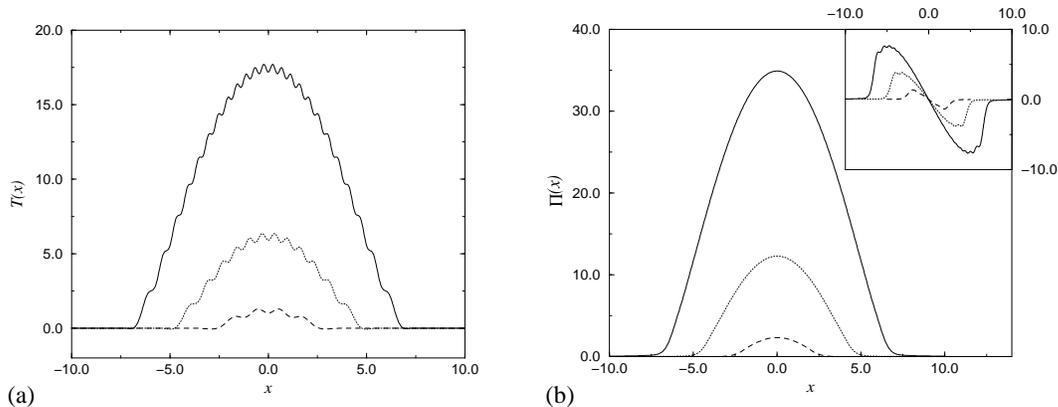,width=0.9\linewidth}}
\caption{Kinetic energy density $T(x)$ (a) and  momentum flux density 
$\Pi(x)$ (b) in units of $\hbar\omega/a_{ho}$ as functions 
of the spatial coordinate $x$ 
for $N$=4 fermions (dashed line), 12 fermions (dotted line) and 24
fermions (solid line). The inset in (b) shows the derivative $d\Pi(x)/dx$
in units of $\hbar\omega/a_{ho}^2$ as a function of $x$ 
(in units of $a_{ho}$). }
\label{fig2}
\end{figure}

The kinetic energy and the momentum flux densities obtained with this
procedure are shown in Fig.~\ref{fig2}
for various number of fermions in 
1D harmonic confinement. Oscillations along the profile and negative tails are
found in the kinetic energy density, while a smooth
positive profile is obtained for the momentum flux density (which
however shows oscillations in its first derivative, see the inset of
Fig.~\ref{fig2}). Other properties of these curves are that the tails
of the momentum flux density profile can be 
expressed analytically in terms of the von Weisz\"acker surface energy
$(\hbar^2/8m)|\nabla n(x)|^2/n(x)$
\cite{march_pra}, and that an approximate form of the kinetic energy
density  can be found using 
the exact density profile in the 
local-density expression $T(x)=\pi^2 \hbar^2[n(x)]^3/6m$ \cite{brack}.

\subsection{Pair distribution function}

Higher order correlations are 
described in the equal-times pair distribution function.
We  focus here on
how  the Pauli exclusion principle is reflected in the pair
distribution function of a gas of $N$ noninteracting fermions in
harmonic confinement.  
The same results 
apply to  a 1D  Bose gas with hard-core interactions, 
since  the boson-fermion mapping holds  for the 
pair distribution function \cite{Girardeau_write}.

The longitudinal pair distribution function  of a quasi-1D Fermi gas is
defined as
\begin{equation}
\rho^{(l)}_2(x_1,x_2)=n(x_1)n(x_2)-F(x_1,x_2),
\end{equation}
where the  function $F(x_1,x_2)$ is given by 
\begin{equation} 
F(x_1,x_2)=\sum_{i,j=1}^N  \phi_i^*(x_1) \phi_j(x_1)
\phi^*_j(x_2) \phi_i(x_2)
\label{effe}
\end{equation}
This function 
can be calculated by an extension of 
the Green's function method 
used in the previous sections to evaluate particle, kinetic energy and momentum
flux densities~\cite{noidiffr}. 
We first rewrite Eq.~(\ref{effe}) in terms of the Green's function
$\hat G(x)$, 
\begin{equation}
F(x_1,x_2)=\frac{1}{\pi^2} \lim_{\varepsilon \rightarrow 0^+}
\sum_{i,j=1}^N {\rm Im} \langle \psi_i | \hat G(x_1) |\psi_j
\rangle \, {\rm Im} \langle \psi_j | \hat G(x_2) |\psi_i
\rangle\;; 
\end{equation}
we then use the property ${\rm Im} A \cdot {\rm Im} B=(1/2)[{\rm Re} (A \cdot
B^*)-{\rm Re} (A \cdot B)]$ to obtain the final expression
\begin{equation}
\fl F(x_1,x_2)=\frac{1}{2\pi^2} \lim_{\varepsilon \rightarrow 0^+} \left\{
{\rm Re} {\rm Tr}\left[ \hat G_N (x_1) \cdot  \hat G^*_N
(x_2)\right] - {\rm Re} {\rm Tr}\left[ \hat G_N (x_1)
\cdot \hat G_N (x_2)\right] \right\}\;,
\end{equation}
where $\hat G_N (x)$ is the first $N\times N$
block of the matrix $\hat G(x)$. 
In the case of harmonic confinement this can be evaluated again by making
use of renormalization techniques and recursive methods:
the evaluation of $F(x_1,x_2)$ is reduced to the
calculation of the determinant of pentadiagonal  $N\times N$ matrices,
which can be factorized into the product
of $2\times2$ matrices.
 
\begin{figure}
\centerline{\psfig{file=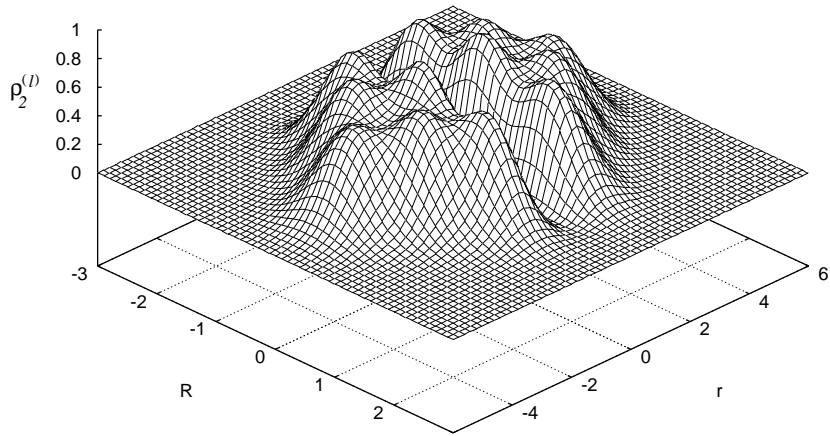,width=0.8\linewidth}}
\caption{Longitudinal contribution $\rho_{2}^{(l)}(r;R)$ 
to the pair distribution function  for $N$=4 fermions 
in quasi-1D  harmonic confinement,  in
units of $a_{ho}^{-2}$,
as a function of the
center-of-mass coordinate $R$  and of the relative
coordinate $r$, both in units of $a_{ho}$. }
\label{fig3D}
\end{figure}

With this Green's function method we have calculated the longitudinal
contribution to the equal-times pair distribution function 
for up to $N=100$ fermions without particular numerical
efforts. Since this function  presents in general a number of maxima of 
order  $N^2$, for the sake of clarity we have shown  in
Fig.~\ref{fig3D}  the full result only for the case of $N=4$ fermions.
We have plotted the pair distribution function                      
in the center-of-mass $R\equiv(x_1+x_2)/2$
and relative  $r\equiv(x_1-x_2)$ coordinates: 
the effect of Pauli exclusion is observable in real space in the $r$
direction as a 
depression of $\rho_{2}^{(l)}(x_1,x_2)$ at short distances.

\section{Dynamical properties}
\label{sec3}

\subsection{Collisionless regime}
The  dynamic structure factor for a non-interacting Fermi gas 
in the collisionless regime is given by
\begin{equation}
\fl S({\bf k},\omega)
=\sum_{i,j}\left|\int d^3 r \, e^{-i {\mathbf  k}\cdot {\mathbf
r}}\phi^*_i({\bf r})\phi_j({\bf r})\right|^2 f(\varepsilon_i) 
[1-f(\varepsilon_j)]\,2 \pi
\delta(\omega-(\varepsilon_j-\varepsilon_i)/\hbar) \;.
\label{s_i_kw}
\end{equation}
We focus here on quasi-1D excitations. The condition in this case is
most stringent than for quasi-1D equilibrium properties (indeed the
Paris experiment \cite{paris_exp}
satisfies the latter condition but not the former):
the transverse excited states
are  not involved in the excitation processes when {\it (i)} the
transverse component $k_\perp$ of the momentum transfer vanishes,
due to orthogonality of
harmonic oscillator wave functions, or {\it (ii)} the energy transfer $\omega$
is smaller than the gap between the chemical potential and the first
transverse excited state. 

\begin{figure}
\centerline{\psfig{file=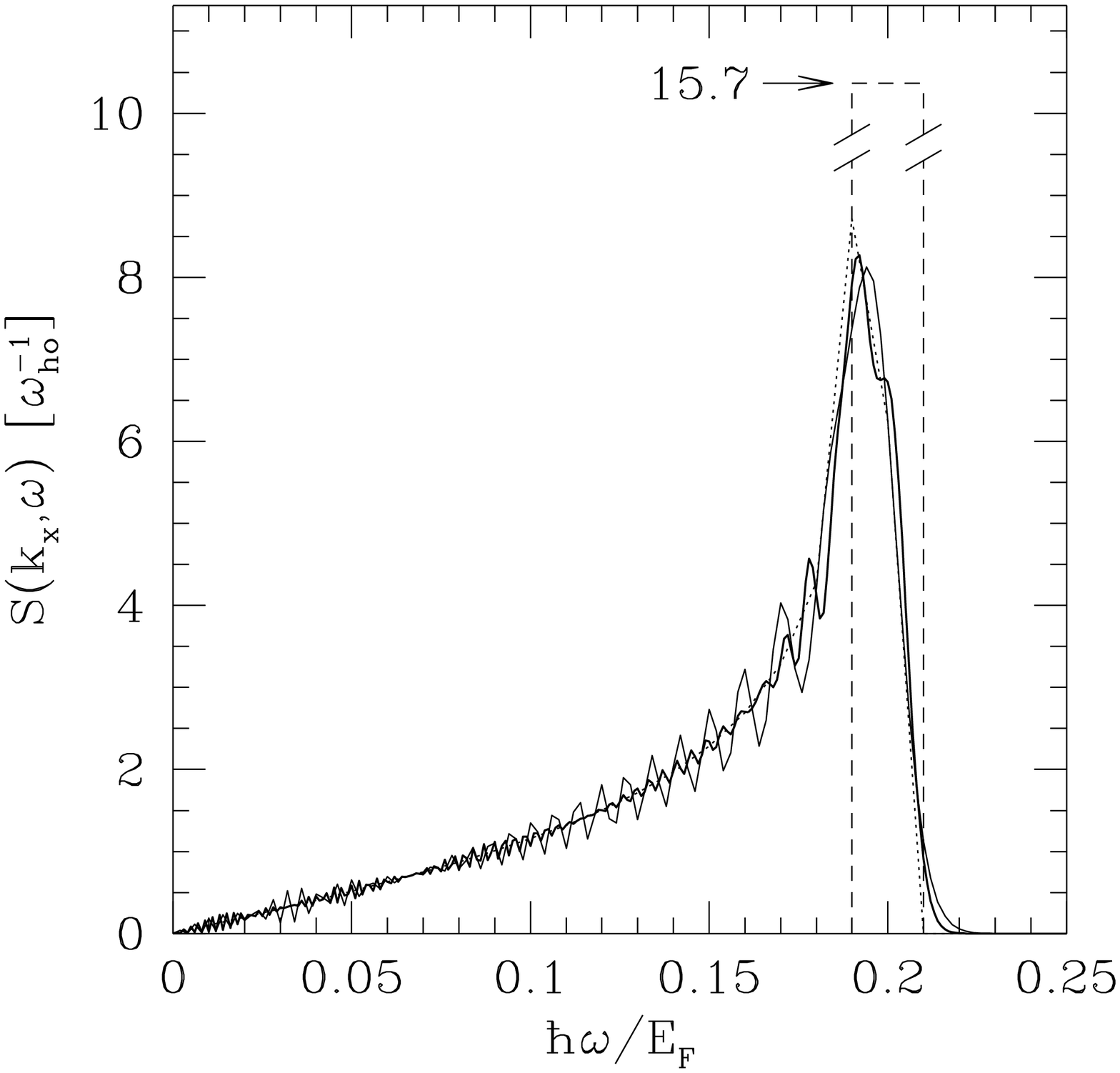,height=0.45\linewidth}
\psfig{file=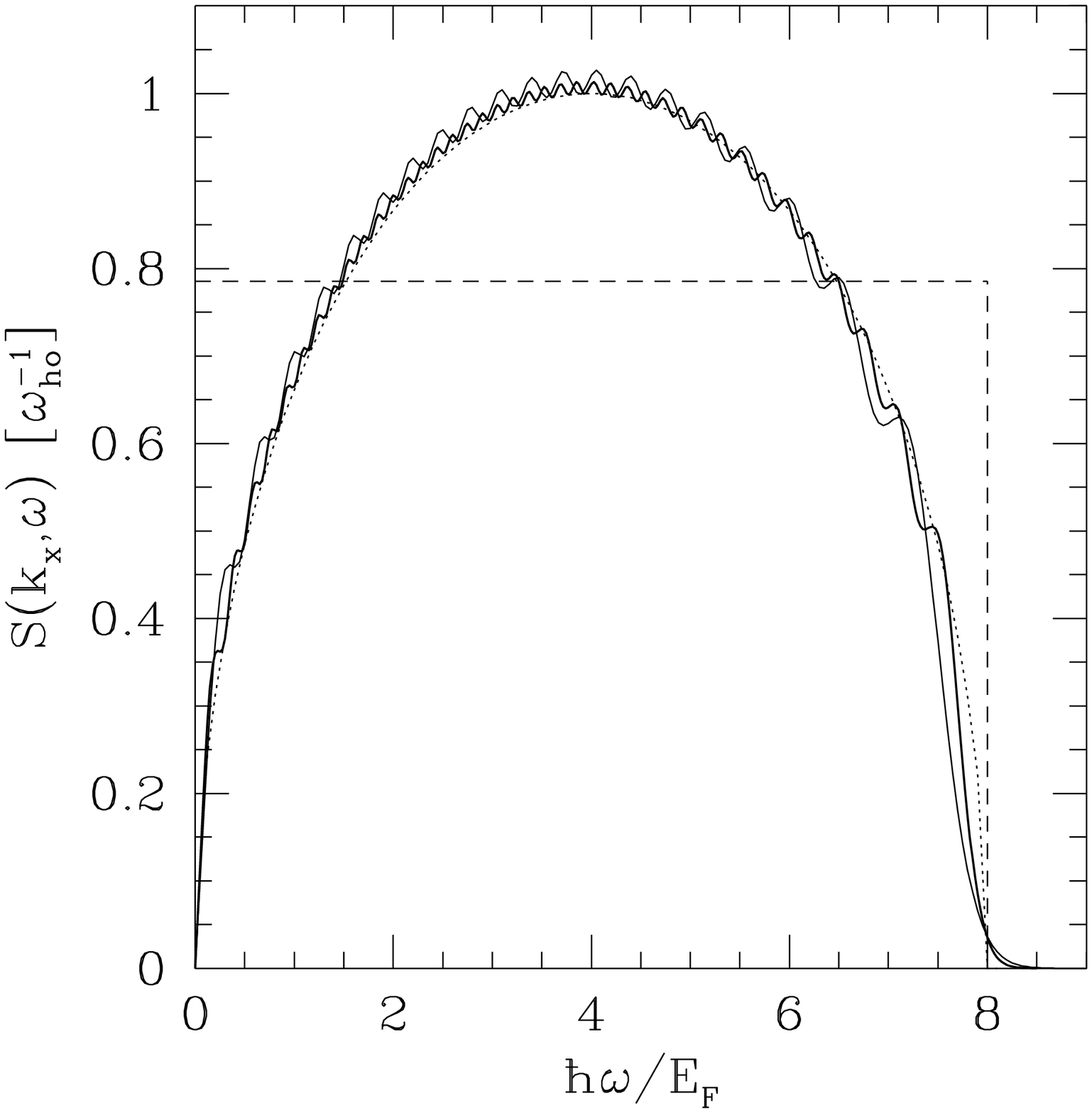,height=0.45\linewidth}}   
\caption{Dynamic structure factor for a 1D ideal Fermi gas in harmonic
confinement (in units of $\omega_{ho}^{-1}$) as a function of
$\hbar\omega/E_F$, where $E_F=N\hbar\omega_{ho}$. Left panel: at
$k=0.1 k_F$ and $N=$500 (solid line) and 100 (bold solid line). Right
panel: at
$k=2 k_F$ and $N=$20 (solid line) and 40 (bold solid line). Both
panels show the LDA spectrum in dotted lines and the homogeneous
result in dashes.}  
\label{fig4}
\end{figure}

In the quasi-1D limit Eq.~(\ref{s_i_kw}) reduces to a one-dimensional
problem and the  transverse-state wave function factorizes out.
By evaluating the overlap integral in Fourier space and 
exploiting  the properties of the Hermite polynomials
 \cite{gradshteyn}, we 
obtain the expression for the dynamic structure factor as \cite{noidiffr}
\begin{equation}
\fl {\cal S}({\bf k}, h)=2 \pi e^{-k_\perp^2/2\lambda} e^{-k_x^2/2}\sum_{i={\rm max}\{N-h,0\}}^{N-1}
\frac{i!}{(i+h)!} \left(\frac{k_x^2}{2}\right)^h 
\left[L_i^h\left(k_x^2/2 \right)\right]^2 \;.
\label{skw_exact}
\end{equation}
Here $h$ is an integer corresponding to a single-atom excitation
of $h$ quanta of the harmonic oscillator,  $L_i^h(x)$ is the $i^{th}$
generalized Laguerre
polynomial of parameter $h$, $\lambda$ is the anisotropy parameter for
the 3D harmonic oscillator and we have set $\hbar=1$, $m=1$ and
$\omega_{ho}=1$.  
The result for the spectrum is shown in Fig.~\ref{fig4} at two
different values of the transferred wave vector $k_x$. 

A good description for the spectrum is already given by the
local-density approximation (LDA), defined as
\begin{equation}
S_{LDA}(k_x,\omega)= \int dx\, n(x) S_{hom}(k_x,\omega;\mu(x))\;. 
\label{skw_lda_def}
\end{equation}
The spatial dependence of the chemical potential $\mu=k_F^2/2=\pi^2n^2/2$
is determined by the
relation $\mu(x)=\mu(n(x))$ and  $n(x)=\sqrt{2N-x^2}/\pi$; and the
expression for the dynamic structure factor of the homogeneous system
is  
$S_{hom}(k_x,\omega)=\pi/k_x k_F$ if
$|\omega_2(k_x)|<\omega<\omega_1(k_x)$ and $S_{hom}(k_x,\omega)=0$
otherwise, with 
$\omega_{1,2}=k_x^2/2\pm k_x k_F$. 
The integral in Eq.~(\ref{skw_lda_def}) can be evaluated analytically
to obtain
\begin{equation}
\fl S_{LDA}(k_x,\omega)=\frac{2}{k_x}\left[\sqrt{2N-\left(\frac{\omega}{k_x}-
\frac{k_x}{2}\right)^2}    
-\sqrt{2N-\left(\frac{\omega}{k_x}+\frac{k_x}{2}\right)^2}\right]\;.
\label{skw_lda}
\end{equation}
From Fig.~\ref{fig4} we see that the LDA description captures the main
features of the spectrum. The presence of the external harmonic
potential significantly modifies the spectrum with respect that of
the homogeneous system. Due to the boson-fermion mapping
\cite{girardeau_td} the spectrum represented in Fig.~\ref{fig4} is
also valid for the Tonks gas.

\subsection{Collisional regime}
We turn now to the long-wavelength limit and we investigate the
analogue of sound-wave propagation. Due to the presence of the
external confinement, the collective excitations are quantized.
We evaluate  here the spectrum and the expression for the density
fluctuations in the linear regime.

The equation of motion for the density profile, as can be derived
exactly from the equations of generalized hydrodynamics
\cite{march_tosi},
 takes the
form
\begin{equation}
m\partial_t^2 n(x,t)=
\partial_x^2 \Pi
(x,t)+\partial_x\left[n(x,t) \partial_x V(x,t)\right]\;.
\label{pinco}
\end{equation}
We define the collisional regime by requiring that  the momentum flux
density $\Pi(x,t)$ depends locally on the 
particle density and the velocity field through the relation
\begin{equation}
\Pi(x,t)=\hbar^2\pi^2 n^3(x,t)/3m+mn(x,t)v^2(x,t)/2\;.
\label{pallino}
\end{equation}
In the linear regime  equations~(\ref{pinco}) and~(\ref{pallino}) 
lead to a closed equation for the density fluctuation $\delta n(x,t)$.
In the homogeneous limit we obtain a phonon dispersion
relation with velocity $c=\hbar k_F/m$.  

In the case of 1D harmonic confinement in order to find a discrete spectrum
we then need to assume a generalized form of
the Thomas-Fermi  approximation (TFA), which reads
\begin{equation}
\delta
\Pi(x,t)\simeq 2\frac{\delta t}{\delta n}\delta n(x,t)=
2[\mu-V_{ext}(x)]\delta n(x,t)
\label{tizio}
\end{equation}
where $t(n)$ is the kinetic energy density and the second equality
follows from the Euler equation for density functional theory. We
assume that Eq.~(\ref{tizio}) holds also outside the classical radius
$X_F=\sqrt{2m\mu/\omega_{ho}^2}$:
this hypothesis is necessary since the solutions of the linearized
Eqs.~(\ref{pinco}) and~(\ref{pallino}) diverge at the classical
boundary as $|x^2-X_F^2|^{-1/2}$ and thus we cannot obtain the
dispersion relation by imposing that the solution should vanish at
$x=X_F$, as in the 3D case \cite{maddalena}.  By matching the
solutions $\delta
n(x) \sqrt{|x^2-X_F^2|}$ inside and outside the classical radius 
we obtain the spectrum \cite{artgp4}  
\begin{equation}
\omega=n \omega_{ho}\;,
\label{spectrum}
\end{equation}
with corresponding  solutions inside the classical radius
\begin{equation}
\delta n_{in}(z)=\frac{\cos(n \,{\rm arccos}(z))}{\sqrt{1-z^2}}
\label{caio}
\end{equation}
and outside the classical radius
\begin{equation}
\delta n_{out}(z)=\frac{\left(|z|-\sqrt{z^2-1}\right)^{n}}
{\sqrt{z^2-1}}\;.
\label{solbout}
\end{equation}
Here we have set $z=x/X_F$.
The divergence of the solutions at $x=X_F$ is unphysical, and
is a consequence of  our approximations: the  linearized TFA
solutions  are not valid around the classical turning points.

\subsection{Non-Linear Schr\"odinger equation}

A possibility to go beyond the TFA is to describe the collisional
Fermi gas through the non-linear Schr\"odinger equation 
\begin{equation}
i\hbar \partial_t \Phi(x,t)=\left[-\frac{\hbar^2}{2m}
\partial_x^2 +V(x,t)+\frac{\pi^2\hbar^2}{2m}
|\Phi(x,t)|^4  \right] \Phi(x,t)\;.
\label{gp4}
\end{equation}
Here, $\Phi(x,t)$ is normalized to the number $N$ of particles in the trap.
According to Kolomeisky {\it et al.}~\cite{kolomeiski}, 
Eq.~(\ref{gp4}) describes the motion of a 1D 
gas of impenetrable bosons in a long-wavelength approach.
Although Eq.~(\ref{gp4}) has been shown to fail in the description of
the dynamics of the phase of the Bose gas \cite{girardeau_td}, 
in the linear regime this equation describes the collisional 
dynamics of the Fermi gas.
Indeed, by setting
$\Phi(x,t)=\sqrt{n(x,t)}\exp[i\phi(x,t)]$ one can transform
Eq.~(\ref{gp4}) into  hydrodynamic  equations for the density $n(x,t)$ and 
the phase $\phi(x,t)$ of the fluid; linearization and neglect of
kinetic energy term $-\partial_x^2 \sqrt{n(x,t)}/2m \sqrt{n(x,t)}$
leads to an expression  which coincides with the linear form
of Eqs.~(\ref{pinco}) and~(\ref{pallino}). 

In the case of a harmonic external potential
a numerical solution for
the low-lying  collective modes  
of  Eq. (\ref{gp4}) 
 has been 
obtained elsewhere. 
The numerical results   are in agreement with
the analytical spectrum 
(\ref{spectrum}) and with the main features of the density
fluctuations  profiles [Eqs.~(\ref{caio}) and (\ref{solbout})]~\cite{artgp4}. 

It is striking that the solution of Eq.~(\ref{gp4}) for the
collective modes in harmonic confinement
always yields the free harmonic-oscillator
spectrum~(\ref{spectrum}), independently of the strength of the
interactions.
In the case of the quadrupole mode, this result can be related  to an
underlying scaling symmetry of the Hamiltonian giving
rise to  Eq.~(\ref{gp4}).
This is the one-dimensional
analogue~\cite{ghosh_tk,ghosh_pk} 
of the hidden symmetry already noticed by Pitaevskii and Rosch for the 2D
Gross-Pitaevskii equation~\cite{pita}:  in the Hamiltonian
the interaction-potential term
\begin{equation}
v(x-x')=\frac{\hbar^2\pi^2}{3m}n(x) \delta(x-x')
\end{equation}
scales under space dilatations $x \rightarrow \lambda x $   
in    the same fashion as the
kinetic-energy term, and this property 
in the case of external harmonic confinement 
leads to a closed equation of motion  for the
expectation value of the quadrupole-mode operator $I=\langle m\hat
x^2/2 \rangle$, 
expressed only in terms of the
total energy $E$:
\begin{equation}
\partial_t^2 I=- 4 \omega_{ho}^2 I + 4 E\,. 
\end{equation}
This yields $\omega=2\omega_{ho}$
for the  frequency of  the quadrupole mode. It
is interesting to compare this value with what is obtained from the 1D
Gross-Pitaevskii equation~\cite{fliesser}, namely
$\omega=\sqrt{5/2}\omega_{ho}$.

\section{Summary and concluding remarks}

\label{sec4}
In summary, we have studied the equilibrium and dynamical
properties of a one-dimensional Fermi gas under harmonic
confinement. We have developed a method which allows to evaluate the
exact equilibrium profiles of the system for up to a large number of
fermions. The density profiles show a shell structure which
is very prominent in this reduced dimensionality. We have calculated the
excitation spectrum of the Fermi gas both in the collisionless regime,
where an analytic exact expression can be obtained, and in the
collisional regime, where we have given a solution in the
generalized Thomas-Fermi approximation. The spectra obtained in the
two regimes coincide --  this is the analogue of the coincidence of the
velocities of zero and first sound in the 1D homogeneous Fermi gas.
The results here presented for the density profiles, pair distribution
function and excitation spectrum describe as well the 1D  Bose gas in
the Tonks regime. The measurement of  excitation frequency of the
quadrupole modes should be a sensible probe of the Tonks-gas phase.

\ack
We thank Professor Tosi for useful discussions and encouragement.
This research is supported by INFM under the project PRA2001.

\end{document}